# T U M

## I N S T I T U T F Ü R I N F O R M A T I K

# Event Handling in ET++

## A Case Study in Algebraic Specification of Object–Oriented Application Frameworks

Klaus Bergner
Bernhard Rumpe

# Event Handling in ET++

## A Case Study in Algebraic Specification of Object–Oriented Application Frameworks*

Klaus Bergner
Bernhard Rumpe
Institut für Informatik, Technische Universität München
Arcisstr. 21, 80290 München, Germany
e-mail: {bergner|rumpe}@informatik.tu-muenchen.de

February 14, 1995

**Abstract**

In this report we investigate the suitability of algebraic specification techniques for the modular specification of complex, object–oriented systems. As an example, part of the event handling mechanism of the application framework ET++ is specified using a variant of the algebraic specification language SPECTRUM.

# Contents

*This work was sponsored by Siemens AG as part of the project "ASE — Automatisierte Software-Entwicklung".

# 1 Introduction

A programmer who wants to use the event handling mechanism of the application framework ET++ in the proper way has two sources of information.

On the one hand, there is the source code, a detailed and accurate description with the disadvantage of being not very readable. This is not only due to the fact that efficient code written in languages like C++ generally tends to be at a low abstraction level, but also due to the fact that information concerning the event handling subsystem is scattered over a lot of different classes. Some information is not explicitly contained in the code: Abstract classes may not provide implementations for some functions and hence may carry no information about the intended use of these "pure virtual" functions.

The other source of information is informal documentation, which is an incomplete and maybe sometimes wrong, but readable and understandable description. It explains not only the intended behaviour of the single objects, but also the behaviour of the whole event handling subsystem which arises from the combination of the interdependent behaviours of its components. Only by reading the informal documentation the programmer can fully understand the rules that should be obeyed in programming with the application framework.

In this paper we report about a study to determine whether algebraic specification may be a suitable formalism for stating desired properties of complex object–oriented systems. This approach seems promising, because the expressive power of algebraic specifications makes it possible to write very abstract and therefore readable specifications that on the other side have a precisely defined semantics and are suitable for theorem proving.

# 2 The Event Handling Mechanism of ET++

ET++ was developed between the years 1987 and 1992 by Erich Gamma and André Weinand [GAM92], [GMW89], [WEI92]. It facilitates especially the development of applications with graphical user interfaces by serving as an object–oriented application framework with hundreds of reusable, interdependent C++ classes. These classes provide basic, generally useful abstractions and mechanisms and can be specialized and adapted in new systems by using inheritance. In this way, a programmer doesn't have to build a totally new program from scratch, but must only write the code specific for the new application.

ET++ offers the programmer a uniform interface to possibly very different underlying window systems (e.g. SunWindows and the X Window System). Only a small set of their features is used, because most of an application's functionality is provided by the reusable classes and the powerful mechanisms of ET++.

A typical ET++ application is structured into subsystems, each consisting of all objects performing a certain task (like drawing windows, handling un- and redoable commands or file management). One of these subsystems is the event handling mechanism, which is responsible for receiving, interpreting, and executing the requests of the user.

For that, the event handling mechanism receives "raw" events (e.g. key presses or mouse clicks) from the underlying window system, finds out, which of the visual objects on the screen (e.g. scrollbars or buttons) is concerned, generates a Command object and directs it to an appropriate ET++ object that can handle the command. In the following, we explain first the concerned classes and then the connection structure of the objects at runtime.

In the inheritance hierarchy of ET++, the classes VObject and Manager are the only direct subclasses of class EvtHandler. Therefore, every EvtHandler object belongs to one of these classes (because of the exclusive use of single inheritance in ET++, an object cannot belong to both classes at the same time). VObject and Manager provide the functionality for event handling via inheriting and adapting the needed operations from class EvtHandler. In this way, all of the various visual objects on the screen (objects of subclasses of VObject, like Button or Window objects), and also all objects managing the data of the application (objects of subclasses of Manager, like Application and Document objects) are capable of performing the needed framework operations on events and commands.

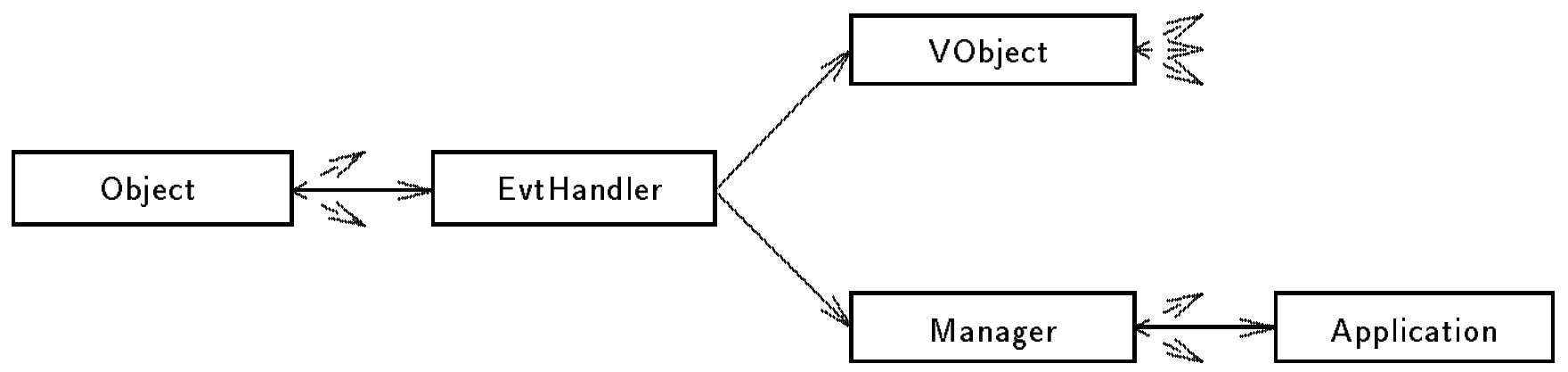

Figure 1: Part of the ET++ Inheritance Hierarchy

At runtime, EvtHandler objects are arranged in the so–called "`part-of`" tree, which is dynamically created and manipulated (figure 2 shows a simple example). The `part-of` tree describes how the event handling objects of the application (which are connected by bidirectional pointers) are nested: A Window could for example contain some Button elements as children and it could itself be contained in a Manager object of class Document. On top of this hierarchy should always be a single object of class Application. All objects above a Manager object must also be of class Manager, whereas all objects beneath a VObject must also be of class VObject, and the first VObject beneath a Manager has to be of class Window.

Whereas the Application object and the other Manager objects themselves are not visible on the screen, every VObject has a screen representation with appropriate coordinates. The bounding box of a visual object that is `part-of` another visual object is geometrically located entirely inside the bounding box of the other object.

At runtime, events (which include a component specifying the screen coordinate to which they pertain) come from the underlying window system and are assigned to the corresponding visual ET++ object of class Window in the `part-of` hierarchy. This assignment mechanism is one of the few parts of ET++ that must be adapted when porting ET++ to a new window system.

From the Window object, events usually traverse the `part-of` tree downwards on a path that consists of visual objects with appropriate coordinates until they have reached a leaf object of the tree (e.g. a Button that contains no further visual objects). In that leaf the events are analyzed: If the object cannot handle an event, it should hand it up to an object higher in the hierarchy, otherwise it usually generates a corresponding Command object and hands this object up. On this chain of event handlers all events should finally reach an event handler that can handle them (that is, generate a Command from them) and all Command objects should finally reach a Manager object where they can be processed.

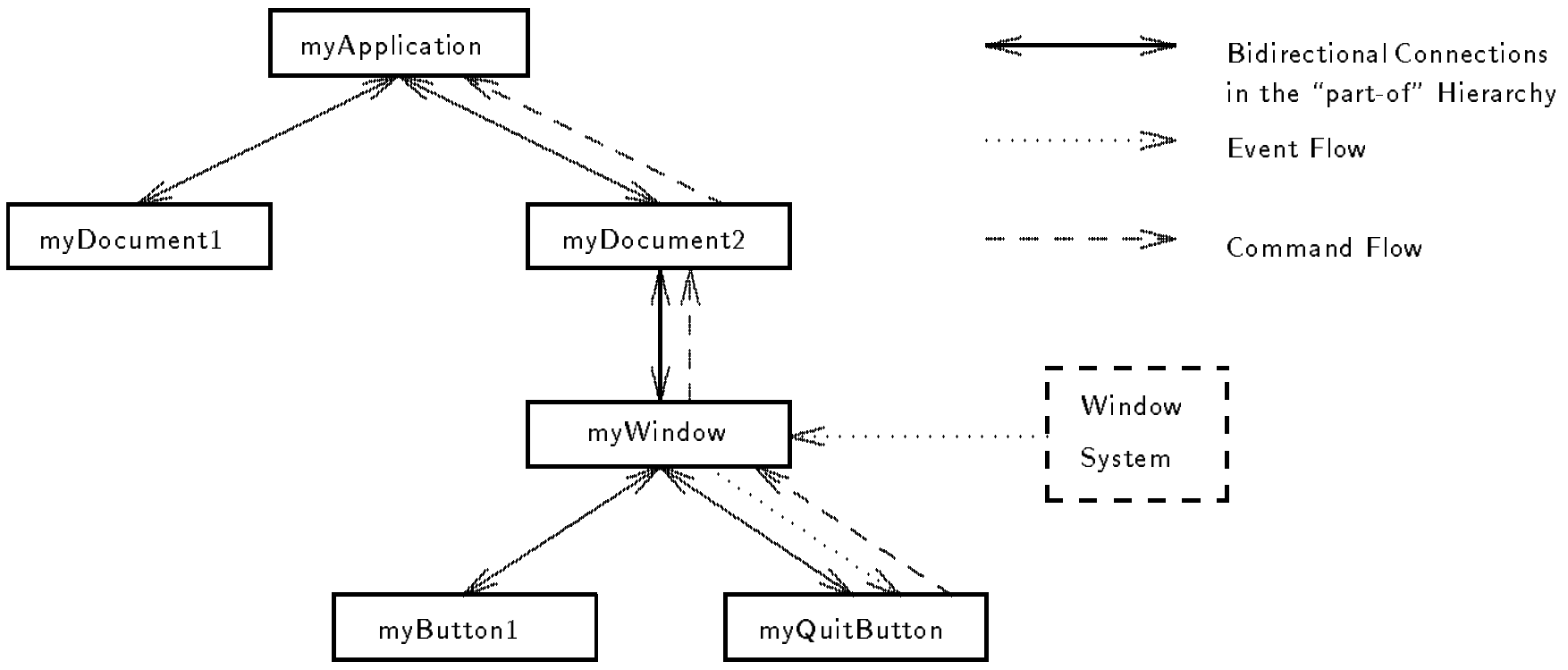

Figure 2: Example of a Simple "`part-of`" Hierarchy at Runtime

In this example, the user of an application has created two documents. One of these has opened a window with two buttons in it. The dotted lines show the event and command flows for a Quit command that terminates the whole application.

While this is the standard behaviour, a programmer is free to change it in various, unpredictable ways: It is for example possible to write an event handling class whose objects don't generate a Command when they get an event of a certain kind, but instead instantly perform the desired action by themselves. Similarly, it is possible (and often required) to change the standard flow of commands at will by creating new upward connections in the `part-of` tree.

In the following, only the (basic) aspects of the event handling mechanism are specified that are intended to be valid in all applications built with ET++; nothing is said about the behaviour of customized event handlers in a special application. In the domain of algebraic specification languages this concept is known as *underspecification*.

# 3 The Specification Language SPECTRUM

There exist already some languages for the specification of object–oriented systems; examples are OOZE [AG92], Troll [EG92], and OS [BRE91].

The specification language we use is a variant of the algebraic specification language SPECTRUM version 0.3 [BFG+92]. SPECTRUM specifications are based on first–order predicate logic and can therefore be understood easily by everyone familiar with that formalism. SPECTRUM allows the specification of partial functions by adding an undefined element $\perp$ to every sort and by offering a definedness predicate $\delta$. The semantics is based on the loose algebra approach, that is, it supports underspecification. Besides the features for specifying datatypes and functions ("specification in the small"), SPECTRUM also offers some operators for combining and transforming specifications ("specification in the large").

We propose a variant of SPECTRUM that is enhanced by some features for the support of object–orientation, like class specifications and an inheritance operator. These additions can be easily mapped to pure SPECTRUM and are intended mainly as notational shortcuts. They are described in section 4.2.

The most important difference between object–oriented languages and algebraic specification languages like SPECTRUM is, that functional specification only deals with stateless values and functions on these values, whereas objects have an identity and a state, which is changed by operations. However, for specification purposes it is possible to abstract away from the internal state: An object in a certain state, characterized by its attribute values (one of which is the immutable object identity), is modeled as a value.

Usually, objects are not isolated: They contain pointers to other objects, such forming systems of cooperating objects. In the framework of algebraic specification, these systems can also be modeled by values. In our case, the whole event handling subsystem of an ET++ application, consisting of many interconnected event handler objects, is modeled as a single value of a tree sort.

The practicability of algebraic specification techniques depends crucially on the existence of encapsulated subsystems with a well–defined functionality. If all objects of an application were interconnected with each other in various ways, one would have to model the state of the whole complex application as a single value — a nearly impossible task. One aspect of this case study is therefore to examine whether existing object–oriented application frameworks contain encapsulated subsystems suitable for algebraic specification.

# 4 The Specification of Event Handling

## 4.1 Basic and Subsidiary Specifications

For the specification of the event handler system we use some basic specifications Nat of natural numbers, SimpleSeq and Seq of sequences, Tree of trees, and PathTree of trees with path handling. These specification units provide reusable components that can be enriched in various ways by adding new axioms, much like abstract C++ classes can be extended by adding new behaviour via providing implementations for virtual functions. The arising hierarchy of specifications is similar to the inheritance hierarchies of object–oriented programming.

We enclose all employed basic specifications with comments in the appendix. It provides a self–contained introduction to SPECTRUM and shows how specification modules can be reused and combined. The element sorts nat, seq, and tree that are introduced in these specifications serve only specification purposes. They are not intended to be implemented by means of C++ classes.

In contrast to this, the subsidiary sorts Id, Token, Command, Point, VObject, Class, and Object correspond with equally named classes: Objects of class Id serve as unique identifiers for event handlers, objects of the classes Token and Command are used for events and commands (which where both explained in section 2), Point objects contain screen coordinates, and VObject is the sort of visual objects. Class provides for every class in the ET++ class hierarchy an object with informations about the class (its objects are used e.g. for querying the name of an object's class at runtime). Finally, Object is the root class of nearly all classes of the ET++ inheritance hierarchy. It provides some basic operations common to most objects in an ET++ application, like the operation .IsA., which tests whether an object is compatible with a certain class (represented by an object of class Class). Although we use these sorts, we do not specify them further, because no knowledge about the internal structure and the behaviour of their elements is needed.

## 4.2 Specification of Class EvtHandler

The class specification EvtHandler (which is included as a whole starting on page 8) is essentially an ordinary specification unit with some syntactic extensions that can easily be translated into standard SPECTRUM notation.

One of these notational shortcuts is, that the statement class before the name of the specification unit ("EvtHandler") automatically introduces an equally named sort EvtHandler. There is an analogy to some programming languages (e.g. Eiffel) where the name of a class is at the same time the name of a module as well as the name of a type [MEY88]. Elements of sort EvtHandler are intended to be implemented by objects of a corresponding C++ class. That means that a specification class X = { ... } (introducing a new sort X that is intended to be implemented by a C++ class X) can also be written as X = { sort X; ... }.

The inherit–operator models the inheritance operator of object–oriented programming languages. The difference to enrich is, that elements of the inheriting class EvtHandler may be used at runtime in place of elements of class Object. Semantically, this means that a subsort relation between the involved sorts is introduced. The specification class Y inherit X = { ... } can therefore also be written as Y = { sort Y; Y $\subseteq$ X; ... }.

As described in section 2, EvtHandler objects are connected via pointers, such yielding a system of interacting objects. In the specification we describe the properties of this system by using an element of sort tree that is renamed to EHSystem. (The axioms for tree elements are given in the specification unit PathTree in the appendix.) Sort w of the value components of EHSystem nodes is set to EvtHandler by renaming it, and the functions getvalue and setvalue are renamed to gethandler and sethandler. Thus, a call of gethandler(ehs,p) yields the event handler that is characterized by the path p from the root of the event handler tree ehs to its corresponding tree node.

Not every tree consisting of event handling objects is a valid EHSystem for an application. To specify this, the operators conforms and validAppEHS are defined. These operators aren't available as functions in the ET++ class EvtHandler — they serve only specification purposes. To indicate this, the keywords op and to are used in their signature.

In the axioms section of conforms, function containsPoint is used. It stems from the class specification VObject (which we haven't included in this paper) and tests whether a screen coordinate of sort Point is contained geometrically in the bounding box of the concerned visual object on the screen. The axioms state that an event with a certain Point coordinate pertaining to a visual event handler object y must also pertain to y's father object (provided the father object is also a graphical object of sort VObject with certain coordinates and not a Manager object) and that events can't pertain to two sons of a single event handler at the same time. On the level of graphical elements on the screen this is equivalent to the fact that every element must be geometrically located entirely inside the bounding box of its father element and that the graphical elements of a certain window can't overlap partly.

In the next axioms section, conforms is employed to specify validAppEHS which can be used to test whether an EHSystem has a valid form and may be legally used in an application. The part (eh IsA Manager $\Leftrightarrow$ $\neg$(eh IsA VObject)) is especially interesting, because it implicitly imposes a constraint on the inheritance structure of sensible applications built with ET++: It makes it impossible for the developer to use objects of newly introduced classes that directly inherit from EvtHandler. The remaining properties have already been explained in section 2: The root of the (nonempty) EHSystem tree should always contain the only object of class Application. All objects above a Manager object must also be of class Manager, whereas all objects beneath an object of class VObject must also be of class VObject. This makes sure that the operator conforms can be applied to each subtree with a root element of sort VObject. Further, the first VObject element directly beneath a Manager has to be of class Window, so that the objects in the subtree beneath this Window have the chance of getting events from the underlying window system.

Another specification section gives the laws for Id–handling: Every event handler must have a unique identifier. The up arrow '↑' in the signature part of the SetId specification

SetId : EvtHandler↑ × Id;

is an abbreviation for the more verbose notation

SetId : EvtHandler × Id → EvtHandler.

For the rest of the operations of class EvtHandler only the signatures can be given. In ET++ these functions are declared virtual, that is, their implementation has to be provided in subclasses or may be changed there. If the programmer wants to assure a special behaviour in some subclasses, that behaviour must be specified in the appropriate class specifications (an example for part of such a specification is given below).

```
class EvtHandler inherit Object = {

 enrich Id + Token + Command + Class + VObject +
          (rename [   tree           to EHSystem,
                      w              to EvtHandler,
                      getvalue       to gethandler,
                      setvalue       to sethandler   ] in PathTree);

 hidden op conforms : EHSystem to bool;

 axioms ∀ v: VObject; s: treeseq; ehs,eht: EHSystem; coord: Point in
   conforms(Θ);
   conforms(mktree(v,s)) ⇔
     (ehs∈s ⇒ conforms(ehs)) ∧
     (¬(v containsPoint coord) ∧ ehs∈s ⇒
          ¬((value ehs) containsPoint coord)) ∧
     (ehs∈s ∧ (value ehs) containsPoint coord ∧
```

```
        eht∈s ∧ (value eht) containsPoint coord ) ⇒ ehs = eht);
endaxioms

op validAppEHS : EHSystem to bool;

axioms ∀ ehs: EHSystem; eh,ehh: EvtHandler; p,q: Path in
  validAppEHS(ehs) ⇔
    (ehs ≠ Θ) ∧
    (eh IsA Manager ⇔ ¬(eh IsA VObject)) ∧
    (eh = gethandler(ehs,p) ∧ p = ≺≻ ⇔ eh IsA Application) ∧
    (eh = gethandler(ehs,p) ∧ ehh = gethandler(ehs,q) ∧ p⊑q ⇔
        (ehh IsA Manager ⇒ eh IsA Manager) ∧
        (eh IsA VObject ⇒ ehh IsA VObject) ∧
        (eh IsA Manager ∧ ehh IsA VObject ⇔ ehh IsA Window)) ∧
    (eh = gethandler(ehs,p) ∧ eh IsA VObject ⇔
        conforms(subtree(ehs,p)));
endaxioms

GetId : EvtHandler → Id;
SetId : EvtHandler↑ × Id;

axioms ∀ ehs: EHSystem; p,q: path; eh: EvtHandler; i: Id in
  p≠q ∧ δ(gethandler(ehs,p)) ∧ δ(gethandler(ehs,q))
          ⇒ GetId(gethandler(ehs,p)) ≠ GetId(gethandler(ehs,q));
  GetId(SetId(eh,i)) = i;
endaxioms

GetNextHandler : EvtHandler → EvtHandler;
FindNextHandlerOfClass : EvtHandler × Class → EvtHandler;
GetMenu : EvtHandler → Menu partial;
DoSetupMenu : EvtHandler↑ × Menu partial;
DoMenuCommand : EvtHandler↑ × MenuCmd → Command partial;
PerformCommand : EvtHandler↑ × Command partial;
SetFirstHandler : EvtHandler↑ × EvtHandler partial;
KbdFocus : EvtHandler↑ × bool partial;
Input : EvtHandler↑ × Point × Token × Clipper partial;
DoIdleCommand : EvtHandler↑ partial;
Send : EvtHandler↑ × Int × Int × Void partial;
Control : EvtHandler↑ × Int × Int × Void partial;
SendDown : EvtHandler↑ × Int × Int × Void partial;
InputKbd : EvtHandler↑ × Token partial;
}
```

In the following, some axioms are given that specify the default behaviour of the functions GetNextHandler and FindNextHandlerOfClass of the EvtHandler class. These functions are used to determine the next event handler that is passed a Command object in case a certain event handler cannot handle that Command. Normally, GetNextHandler(eh) yields eh's father in the `part-of` tree, whereas FindNextHandlerOfClass(eh,cl) yields eh's first ancestor in the path from eh to the root of the tree that is compatible with class cl.

If in a given application this default behaviour is assured for every object that is compatible with class EvtHandler (or with a heir class of EvtHandler, respectively), these axioms can be inserted into the specification of EvtHandler (or into the specification of the heir class, respectively), thus fully specifying the previously underspecified behaviour.

$$
\begin{array}{l}
\mathsf{axioms}\ \forall\ \mathsf{eh,ehh: EvtHandler;\ cl: Class;\ p,q,r: Path: ehs: EHSystem\ in} \\
\quad \mathsf{eh} = \mathsf{gethandler(ehs,p)} \wedge \neg(\delta(\mathsf{GetNextHandler(eh)})) \Leftrightarrow \mathsf{p} = \prec\succ; \\
\quad \mathsf{eh} = \mathsf{gethandler(ehs,p)} \wedge \mathsf{ehh} = \mathsf{gethandler(ehs,q)} \Rightarrow \\
\qquad (\mathsf{GetNextHandler(ehh)} = \mathsf{eh} \Leftrightarrow \mathsf{p} = \mathsf{lead(q)}) \wedge \\
\qquad (\mathsf{FindNextHandlerOfClass(ehh,cl)} = \mathsf{eh} \Rightarrow \\
\qquad\quad \mathsf{eh\ IsA\ cl} \wedge \mathsf{p} \sqsubseteq \mathsf{q} \wedge \\
\qquad\qquad \forall \mathsf{r.p} \sqsubset \mathsf{r} \sqsubset \mathsf{q} \Rightarrow \neg(\mathsf{gethandler(ehs,r)\ IsA\ cl})); \\
\mathsf{endaxioms}
\end{array}
$$

# 5 Conclusion

This study has shown that the algebraic specification of complex, object–oriented application frameworks can have some advantages, but also bears a number of difficulties.

There is no doubt that a formalism for the succinct and clear description of object–oriented frameworks is urgently needed. It could help to cure the perhaps biggest disadvantage of the framework approach: the difficulty of understanding how to use and to adapt the various classes and mechanisms of a complex framework in the intended way.

The main reason for this difficulty is, that the informations for a certain mechanism are usually scattered over a lot of different places in the source code of a framework. In our case, not only the classes EvtHandler, Token, and Command had to be examined in detail, but also the Manager, VObject, and Window classes. In general, this means that for the specification of a superclass, the source code of all subclasses have to be examined, too. Only then it can be avoided to state "axioms" in the specification of a superclass that are violated by objects of one of its subclasses.

Some of the information about the framework can not be found in the source-code at all: To understand the intended properties of virtual functions, for which no or only a simple default implementation is given, the documentation must be read. In our case, the intention of some mechanisms wasn't described in the documentation at all and an expert had to be consulted. A good example is the Id–handling. It only makes sense, if every event handler has its own, unique Id. Because the programmer is responsible for setting Ids, he or she could also decide to implement a mechanism where a number of event handlers may have the same Id. Whereas the source code and even the documentation don't forbid that, our specification does.

The possibility to change the behaviour of the event handling mechanism quite drastically is an intended feature of ET++: It implies adaptibility to many problems. On the other side, it also causes some disadvantages. First, it makes the comprehending of applications more difficult, because every programmer may freely modify the mechanisms of the application framework in a highly non–standard way. It also makes it impossible to give a complete, formal description of all aspects of the ET++ event handling system: If the programmer is legally allowed (and even encouraged) to change certain aspects of a system at will, no general axioms concerning these aspects can be given. In our case, only the basic rules concerning the behaviour and structure of the event handling mechanism could be given; most of the essential functions had to be left unspecified, because the programmer is free to modify their standard behaviour at will.

A problematic issue with functional, algebraic specification languages is, that they are no practical tools for specifying applications whose objects are interconnected in various ways by means of pointers. In this case, one would have to model the state of the whole application as a single value, resulting in incomprehensible and therefore useless specifications. Though most of the objects in an ET++ application are interacting and therefore interconnected event handlers, they are always organized in a simple, tree–like structure. As we have shown, it is possible to describe their behaviour and connections very succinctly and clearly. However, a programmer who adds code to the framework could in principle add new connections between random event handlers in the tree (cf. section 2), thus making it very difficult to fully specify the processing of events. However, it is our conjecture that in a well–designed, comprehensible object–oriented system the communication between objects is always structured in a very regular way that is suited for functional specification.

## Acknowledgements

We want to thank Peter Sommerlad from Siemens, Hans J. Fröhlich, Rainer Weinreich, and Reinhold Plösch from Johannes-Kepler-Universität, and Dieter Nazareth and Barbara Paech from Technische Universität München for providing useful comments on draft versions of this paper.

# 6 Bibliography

## References

[AG92] Antonio J. Alencar and Joseph A. Goguen. *OOZE with Examples.* Technical report, Programming Research Group, Oxford University Computing Laboratory, 1992.

[BFG+92] M. Broy, C. Facchi, R. Grosu, R. Hettler, H. Hußmann, D. Nazareth, F. Regensburger, and K. Stølen. *The Requirement and Design Specification Language* SPECTRUM*, An Informal Introduction, Version 0.3.* Technische Universität München, Interner Bericht TUM-I9140, 1992.

[BRE91] Ruth Breu. *Algebraic Specification Techniques in Object Oriented Programming Environments.* Springer-Verlag, 1991.

[EG92] Hans-Dieter Ehrich and Martin Gogolla. *Objects and Their Specification.* In M. Bidoit, C. Choppy, H. Ehrig, F. Orejas, and H. Reichel (editors), *8th Workshop on Abstract Data Types*, LNCS. Springer-Verlag, 1992.

[GAM92] Erich Gamma. *Objektorientierte Software–Entwicklung am Beispiel von ET++.* Springer-Verlag, 1992.

[GMW89] Erich Gamma, André Weinand, and Rudolf Marty. *Design and Implementation of ET++, a Seamless Object–Oriented Application Framework.* Structured Programming, Vol. 10, No. 2. Springer-Verlag, 1989.

[MEY88] Bertrand Meyer. *Object–Oriented Software Construction.* Prentice Hall, 1988.

[WEI92] André Weinand. *Objektorientierte Architektur für graphische Benutzungsoberflächen.* Springer-Verlag, 1992.

# A Appendix: Basic Specifications

## A.1 Natural Numbers

The following specification with the name Nat consists essentially of two parts:

First comes a signature part, where sorts (here only sort nat) and functions with their functionality are introduced. In our case, the constant functions 0, 1 and 2, the successor and predecessor functions succ and pred, and infix functions for addition .+. and test for $.\leq.$ are present. The double–headed arrow denotes strict and total functions, whereas the keywords prio and left describe the binding power and associativity of the respective operators.

In the second part, axioms for the natural numbers are given. The natural numbers can be inductively generated by repeated application of the successor function succ to the constant 0 and certain axioms of first–order predicate logic are valid. The operator $\delta$ denotes a definedness predicate.

```
Nat = {

  sort nat;

  0,1,2 : nat;
  succ : nat ↠ nat;
  pred : nat → nat strict;
  .≤. : nat × nat ↠ bool prio 5;
  .+. : nat × nat ↠ nat prio 6: left;

  nat generated by 0, succ;

  axioms ∀ a,b: Nat in
    1 = succ(0); 2 = succ(1);
    a≠b ⇒ succ(a)≠succ(b); succ(a) ≠ 0;
    ¬(δ(pred 0)); pred(succ a) = a;
    a + 0 = a; a + succ(b) = succ(a+b);
    a ≤ a+b; ¬(a+succ(b) ≤ a);
  endaxioms
}
```

## A.2 Simple Sequences

Sequences with the generic element sort w are specified. They can be generated by repeated application of the append function to the empty sequence $\prec\succ$. Functions for selecting the first element of a sequence and the rest of the sequence are available.

Other operations are the constructor $\prec.\succ$ for building one–element sequences, the functions lead and stock that behave similar to rest and append, only at the end of the sequence, and the function .^. for concatenation of two sequences.

```
SimpleSeq = {

  sort w , seq;

  data seq = ≺≻ | append( first: w, rest: seq );

  ≺.≻ : w ↠ seq;
  stock : seq × w ↠ seq;
```

```
  lead : seq → seq strict;
  .^. : seq × seq ↠ seq prio: 6 left;

  axioms ∀ a,b: w; s: seq in
    ≺a≻ = append(a,≺≻);
    stock(≺≻,a) = ≺a≻; stock(append(b,s),a) = append(b,stock(s,a));
    lead(stock(s,a)) = s; ¬(δ(lead(≺≻));
    s^append(a,t) = stock(s,a)^t; s^≺≻ = s;
  endaxioms
}
```

The data–notation is an abbreviation for the following signature

```
  ≺≻ : seq;
  append :  w × seq ↠ seq;
  first : seq → w strict;
  rest : seq → seq strict;
```

in combination with the axioms

```
  seq generated by ≺≻, append;
  axioms ∀ a: w; s: seq in
    first(append(a,s)) = a; ¬(δ(first ≺≻));
    rest(append(a,s)) = s; ¬(δ(rest ≺≻));
  endaxioms.
```

These axioms imply the initiality of the sequence datatype, that is, one can deduce

```
  ≺≻ ≠ append(a,s);
  a≠b ∨ s≠t ⇒ append(a,s) ≠ append(b,t).
```

## A.3 Sequences

The specification of simple sequences is enriched by a function that gives the length of a sequence (length), mixfix functions that select the nth element (.[.]) and a finite subsequence (.[.,.]), and infix functions that test for inclusion of an element (.∈.) and whether sequences are prefixes of other sequences (.⊑. and .⊏.).

The application of the enrich operator on the specification units SimpleSeq and Nat makes the signatures and axioms of these two units available.

```
Seq = { enrich SimpleSeq + Nat;

  length : seq ↠ nat;
```

```
  .[.] : seq × nat → w strict;
  .[.,.] : seq × nat × nat ↠ seq;

  axioms ∀ a: w; m,n: nat; s,t: seq in
    length≺≻ = 0; length(s^t) = length(s)+length(t);
    ¬(δ(≺≻[n])); ¬(δ(s[0]));
    append(a,s)[1] = a; append(a,s)[2+n] = s[1+n];
    ≺≻[n,m] = ≺≻; append(a,s)[1,2+m] = append(a,s[1,1+m]);
    append(a,s)[2+n,1+m] = s[1+n,m];
    (n≤m) ∧ (n≠m) ⇒ s[m,n] = ≺≻; s[0,n] = s[1,n];
  endaxioms;

  .∈. : w × seq ↠ bool;
  .⊑. : seq × seq ↠ bool;
  .⊏. : seq × seq ↠ bool;

  axioms ∀ a,b: w; s,t: seq in
    ¬(a∈≺≻); a∈≺b≻ ⇔ a=b; a∈s^t ⇔ a∈s ∨ a∈t;
    s ⊑ t ⇔ (t≠≺≻ ∧ s[1]=t[1] ∧ rest(s)⊑rest(t)) ∨ s=≺≻;
    s ⊏ t ⇔ s⊑t ∧ s≠t;
  endaxioms;
}
```

## A.4 Trees

Ordered trees with an unbounded number of sons for each node are specified. A tree node of a non–empty tree consists of a value part of the generic parameter sort w and a sequence sonseq of the son–trees. From these two components a tree is built via the constructor mktree. The empty tree is denoted by Θ.

The rename operator changes the names of the sorts and functions in specification Seq according to the given renamelist. Note that w in specification Tree references to two different sorts: the w in the renamelist is renamed to tree, in this way instantiating the generic sort parameter w in Seq, whereas the w below is a freshly introduced generic parameter sort for the elements of the tree.

```
Tree = { enrich (rename [ w to tree, seq to treeseq ] in Seq);

   sort w;

   data tree = Θ | mktree( value: w, sonseq: treeseq );
}
```

## A.5 Trees with Paths

The above specification of trees is enriched, yielding trees with support for easy manipulation of the contents of single nodes (via the functions getvalue and setvalue) and access to whole subtrees (via function subtree). Locations in trees are specified by paths, which in this context are sequences of natural numbers identifying single nodes in a tree. The sequence append(3,≺2≻) would for example identify the second son of the third son of the root of a tree.

```
PathTree = { enrich Tree + (rename[ w to nat, seq to path ] in Seq);

  getValue : tree × path → w strict;
  setValue : tree × path × w → tree strict;
  subtree : tree × path → tree strict;

  axioms ∀ t: tree; a,b: w; p: path; n: nat in
    getValue(t,≺≻) = value(t);
    getValue(t,≺n≻^p) = getValue(sonseq(t)[n],p);

    ¬(δ(setValue(Θ,p,a)));
    setValue(mktree(b,s),≺≻,a) = mktree(a,s);
    setValue(mktree(b,s), ≺n≻^p, a) =
            mktree(b, s[1,n−1] ^ setValue(s[n],p,a) ^ s[n+1,length(s)]);

    subtree(t,≺≻) = t;
    subtree(t,≺n≻^p) = subtree(sonseq(t)[n],p);
  endaxioms
}
```